\documentclass[10pt,conference]{IEEEtran}
\usepackage{graphicx}
\usepackage{hyperref}
\usepackage{booktabs}
\usepackage{array}
\usepackage{tabularx}
\usepackage{threeparttable}
\usepackage[table]{xcolor}
\usepackage{cite}

\title{Optimizing Workflow for Elite Developers: Perspectives on Leveraging SE Bots}
\author{Zhendong Wang}
\author{
        Zhendong Wang$^{\dagger}$, Yi Wang$^\ddagger$, David Redmiles$^\dagger$ \\
    $^\dagger$University of California, Irvine, USA\\
    $^\ddagger$Beijing University of Posts and Telecommunications, China\\
    Email: zhendow@uci.edu, wang@cocolabs.org, redmiles@ics.uci.edu
}

\begin{document}

\maketitle
\begin{abstract}
Small-scale automation services in Software Engineering, known as SE Bots, have gradually infiltrated every aspect of daily software development with the goal of enhancing productivity and well-being. While leading the OSS development, elite developers have often burned out from holistic responsibilities in projects and looked for automation support. Building on prior research in BotSE and our interviews with elite developers, this paper discusses how to design and implement SE bots that integrate into the workflows of elite developers and meet their expectations. We present six main design guidelines for implementing SE bots for elite developers, based on their concerns about noise, security, simplicity, and other factors. Additionally, we discuss the future directions of SE bots, especially in supporting elite developers' increasing workload due to rising demands.
\end{abstract}

\section{Introduction}

With the advancement of automation technologies, the field of software development has been compelled to keep pace with the progression of automation. Various automation services, commonly known as SE bots, have gradually infiltrated all aspects of software development, with a unified goal of enhancing productivity~\cite{wessel2018power, storey2016disrupting}. These SE bots assist in setting up CI/CD pipelines~\cite{wang2022specialized}, enabling bot-assisted workflows~\cite{wessel2020effects, moharil2022between}, and more recently, providing AI programming assistance in low-level source code production~\cite{robe2022pair, chen2021evaluating}. 

While each of the above domains is important in automating software engineering, SE bots have become ubiquitous in the field and have assisted with various aspects of daily technical and non-technical tasks~\cite{wang2022specialized}. Previous studies have demonstrated the significant role of \textit{elite developers} and their non-technical tasks in developing software and maintaining OSS projects, and how these tasks are associated with productivity and other technical outcomes~\cite{mockus2002two,trinkenreich2020hidden, wang2020Unveiling}. Therefore, it is necessary to provide a holistic view of how the latest SE bots assist in these dimensions of software engineering activities and to fully utilize and improve the designs of these SE bots.

This study aims to provide a comprehensive understanding of the current bot-assisted workflow, as perceived by elite developers, building upon prior research and practice in SE bots. Specifically, the goal is to explore practitioners' experiences of using SE bots and their expectations of bot development. Prior studies have highlighted the widespread use of various bots in OSS development, particularly in socially successful and contribution-intensive projects~\cite{wessel2018power, wang2022specialized}. However, little is known about how elite developers incorporate these bots into their workflow and the perceived value they bring.


This study is divided into two parts: First, 
the study conducts semi-structured interviews to collect feedback on expectations for future SE bots and how these bots may fit into developers' workflow. The interview data is analyzed through an open coding procedure by annotating raw interview transcripts and conducting thematic analyses to extract in-depth findings. Second, by synthesizing interview results with prior research, this study presents six main design guidelines for improving and designing SE bots for elite developers, based on their concerns about noise, usability, security, and other factors. 

The findings of this study suggest that software engineering (SE) bots have been immensely helpful to elite developers in both technical and non-technical activities. Nevertheless, participants reported that there is still room for improvement in terms of usability. The study therefore offers recommendations on how to allocate limited engineering and research resources to improve, with a focus on enhancing developers' workflows. Furthermore, the study highlights the potential for future SE bots to better support elite developers in managing their increasing workload resulting from rising demands.

\section{Background}
The prevalence of automation technology in the form of SE bots has made them a crucial component of daily engineering processes within many OSS organizations, serving as the primary interface for human-AI interactions~\cite{liu2020understanding}. They serve as an extension of the development team, i.e., repository \textit{butler}, providing automated feedback and monitoring of code quality, release management, and community engagement~\cite{lebeuf2018software, wang2022specialized}. With their ability to perform repetitive tasks and improve efficiency, SE bots have enabled developers to focus on more creative and complex tasks. Overall, SE bots have significantly impacted the way OSS organizations operate, making their daily engineering processes more streamlined~\cite{erlenhov2019current}.

Despite the widespread adoption of SE bots in OSS, there are still challenges and limitations that need to be addressed. Previous studies have employed various methods, such as surveys, interviews, and empirical analyses, to identify these issues and propose solutions. Frist, one major issue is the limited understanding of user needs and preferences, which has resulted in significant usability problems~\cite{wessel2022bots, liu2020understanding, erlenhov2020empirical}. Second, without clear standardization, practitioners have difficulties trusting automated decision-making processes~\cite{erlenhov2020empirical, brown2019sorry}. Third, the evolving nature of software technology poses challenges in maintaining and updating bots, and they are often slow to keep up~\cite{wang2022specialized, erlenhov2019current}. To address these issues, several bot development frameworks have been introduced with the aim of providing a standardized and transparent bot creation process, including OpenBot, Probot, Octokit, and others. Additionally, Wessel has provided general guidelines for developing bots for GitHub, with a focus on supporting social coding platforms~\cite{wessel2022guidelines}.


\section{Interview for Engineering Experience}
The study employed a purposive sampling technique and a semi-structured approach to conduct interviews with participants selected from a list of OSS projects with bot deployment~\cite{wang2022specialized}. Participants who were categorized as elite developers were invited through email or text messages sent to public channels of their respective organizations~\cite{wang2020Unveiling}. The interview process consisted of 10 pre-defined questions grouped into three main sections, namely introduction, workflow, and expectations of SE bots (see Tab. \ref{tab:interview_participants}). All interviews, which lasted between 25 to 45 minutes each, were conducted across 2021 and 2022. Participants' responses were analyzed using an open coding procedure, and the raw interview transcripts were annotated for thematic analysis to extract meaningful findings~\cite[Ch.4]{braun2012thematic}. To ensure coding schema validity, two researchers assessed one participant's transcript, and substantial agreement was achieved (Cohen's $\kappa = 0.66$). The study's primary objectives were to determine the workflow and expectations of SE bots, which were the major findings of the research. We elaborate on the major emergent themes in the subsequent sections.


\begin{table}[t]
\centering
\small
\caption{Demographic details of interview participants}\label{tab:interview_participants}
\begin{tabular}{@{}lllll@{}}
\toprule
ID & Gender & Country & Most Recent FTE Role     \\ \midrule
P1 & Male   & Canada  &  Software Engineer                    \\
P2 & Male   & Canada  &  Support Engineer                     \\
P3 & Male   & USA     &  Software Engineer                                      \\
P4 & Female & USA     &  Data Engineer                    \\
P5 & Male   & USA     &  Sr. Software Engineer                      \\
P6 & Male   & USA     &  Research Engineer               \\
P7 & Male   & USA     &  Sr. Software Engineer           \\ \bottomrule
\end{tabular}
\end{table}

\subsection{Interview Findings}

\subsubsection{Workflow with SE Bots}
The results of the interviews have demonstrated that automation has a significant impact on the daily workflow of developers. Technical SE bots have increasingly integrated into the repository, affecting several aspects of the engineering process, including test executions, deployment, and other development pipeline stages. The integration of SE bots into a repository's workflow and maintenance takes two main approaches, depending on the triggering actors: \textit{User-based} and \textit{System-based}~\cite{erlenhov2019current, lebeuf2019defining}. 

\textit{User-based} bots support and enhance human tasks, such as the pull request workflow~\cite{wessel2022guidelines}. Pull requests are widely accepted as the basic unit of open-source technical contribution. In well-maintained and automation-assisted OSS repositories, initiating a pull request involves following a predefined format by filling in several fields. External contributors submit pull requests, and two automation services may intervene: pull request formatting and CLA collectors. Pull request formatting checks whether the target pull request follows a specific format, and the Contributor License Agreement (CLA) bot requests contributors' signatures on the agreement, such as releasing intellectual property if they have not already done so (P1 and P2). Some repositories provide visualization or text-based summaries to assist by leaving bot-generated comments below that pull request (P3). If these checks pass, the pull request proceeds to build and test. SE bots such as \texttt{codecov} and \texttt{coverall} build and generate a test report as additional review information, which helps the code review process. Some repositories use a pull request triage bot to label and assign code reviewers to manage their backlog (P6). Finally, deployment previews provided by bots such as \texttt{netlify} and \texttt{Travis} assist code reviewers in assessing the quality of the implementation in the pull requests (P1). Other bots proactively help developers' daily workflow, including auto-generating release drafts based on pull requests and commit messages, and welcoming first-timers for submitting issues.

\textit{System-based} bots monitor the artifacts of repositories, update resources from the internet, and remind or provide actionable measures to correct unwanted human behaviors. One of the most popular SE bots in this category is \texttt{depandabot}. According to positive feedback from developers (P1, P3, and P5), the \texttt{depandabot} provides significant technical value. The \texttt{depandabot} scans source code headers and package dependencies in the repository. When there is a critical release or update of these packages, the \texttt{depandabot} initiates a pull request to update the source code. Although the bot is occasionally ``unnecessarily sensitive'' (P3), this feature helps developers maintain repository dependencies up-to-date and mitigate security concerns. Another active example is the \texttt{stale} bot (P1, P3, and P4). This bot monitors all repository issues and pull requests. When these items have not received any actions (including comments and events) for a certain period, the bot labels them and/or comments with a ``stale'' indicator to remind developers to take actions with inactive backlog items or requests. This bot and its similar alternatives (\texttt{marypoppins} bot, etc.) helped developers manage repository backlogs, especially with a growing community (P2). Other system-based bots were also mentioned in developers' daily workflow, for instance, activity summary (P7) and community acknowledgment (P2 and P5).

\subsubsection{Notes on Platform Notifications}
One notable finding is that developers heavily rely on GitHub's notification system to stay up-to-date with the latest changes in their artifacts and contributor/user community. As elite developers of a repository, they receive various types of notifications by default, such as being assigned to an issue or pull request, opening a pull request, issue, or team discussion post, commenting on any of these threads, subscribing to a thread, changing the state of a thread (issue events), and being mentioned with ``@username.'' These notifications offer a comprehensive view of both technical and social updates of a repository. Elite developers often use notifications to provide timely support to the community, as quick responses to their contributors and users are seen as essential to maintaining positive relationships with users and enhancing reputation and publicity (P1 and P5).

For instance, P2 emphasized the importance of supporting the community and stated that ``\textit{you often} [need to] \textit{go to the Slack channel and Google Groups, and that's like a really active community. So you're the person in a way going through the best you can and interacting with community members.}'' Similarly, as a founder of their software project and pushing it into its commercial path, P1 mentioned, ``[Having this user community as an asset], \textit{so we want to make sure, the community likes what I was changing, and what got developed, right?}'' Although many OSS participants are full-time employees elsewhere, and sometimes do not have full effort in planning and managing their projects, the community's requests have presented both advantageous and disadvantageous effects on the elite developers. On the one hand, elite developers have a backlog of community requests to work on (P1, P5, P6), but on the other hand, they might feel stressed when the latest requests come in (P1, P2, and P5). Therefore, checking notifications is a high-priority task in developers' workflow, but it can also be a stressful experience.

\subsubsection{Values Brought by SE Bots}

In terms of the benefits of SE bots, there are three distinct advantages. Firstly, these bots provide significant technical value to developers, as identified by a number of participants (P1, P2, P3, P5, and P7). As previously noted, SE bots automate various aspects of developers' workflows in both user- and system-based ways. Bots that are active in the CI/CD process save human effort in building and testing, and provide additional information to evaluate the code review process. Participant 5 stated, ``...\textit{One, two, three...I guess we now have four checks in CI now}, [and] \textit{I don't think I would remember to run them all every time} [not if the bot would run them for me],'' highlighting that automated workflows help remind or directly execute checks for developers in this rapidly evolving development environment. This assistance greatly benefits them, in their words, ``\textit{provided technical values} (P6)'' to their daily development workflow, i.e., significantly improve their productivity working with the CI/CD pipeline and enhance merged codebase quality via various automated quality assurance measures.


Secondly, SE bots provide timely support for the community, according to our study's participants. As elite developers often have to dedicate significant effort to support and organize their community, automation support for these tasks can be helpful. SE bots can assist in verifying whether contributors have signed CLAs, welcoming first-time contributors, and enforcing repository guidelines and code of conduct, among other administrative tasks. This type of assistance serves as a buffer between developers' community support work and other activities, allowing elite developers to focus on creative work or other aspects of their lives. For instance, one participant noted that not having to monitor community updates constantly would be beneficial: ``...\textit{not having to watch my phone's notification would feel good}. (P1)''

Finally,  our study found that SE bots also help developers alleviate their mental load by handling various engineering tasks. OSS elite developers often worry about their project's various responsibilities and feel the need to take care of everything in their project. As one participant remarked, contributing to OSS is not like their daytime work, and they must handle all aspects of the project themselves: ``...[\textit{contributing to} OSS] \textit{not like what I work during the daytime, I need to take care of everything in this project.} (P6)'' 

\subsubsection{Challenges of Applying SE Bots}

As per previous studies on SE bots~\cite{wessel2018power, liu2020understanding}, our study identified two significant challenges associated with using SE bots: excessive notifications and limited interactivity. The first challenge that was the increased number of notifications, particularly from proactive system-based bots during irregular development times. Unlike reactive bots that provide immediate feedback during work sessions of CI/CD pipelines, proactive bots such as \texttt{stale} and \texttt{weekly-digest} may post new threads beyond core development hours. Since participants in our study, such as P1 and P5, used a phone app to monitor the repository, additional notifications could annoy and stress developers.

The second major challenge was the limited interactivity of SE bots. As mentioned in a previous study~\cite{wang2022specialized}, popular SE bots typically employ rule-based design, leading to pre-defined reactions to specific repository events or thread content. While these restricted ways of interaction can be effective and direct for experienced developers who require additional information in CI/CD pipelines, the lack of interactivity can be problematic when working with novices and community users: ``\textit{I think the bots that we've used haven't been interactive. I think they've all just been kind of dump}[\textit{ing}] \textit{a lot of information.} (P3)'' As a community support engineer, P2 noted that due to the lack of interactivity, there were no quick automated ways to ensure that the community knew what was happening with the project or how they felt about it.
Compared with more technical CI/CD bots, community support bots require certain conversational capabilities to provide contributor guidelines.

\begin{table}[t]
\center
\scriptsize
\caption{Evidence Supporting DGs of SE Bots for Elite Developers}\label{tab:dg}
\newcolumntype{Y}{>{\raggedright\arraybackslash}X}
\centering
\begin{tabularx}{\columnwidth}{p{0.03\columnwidth} Y p{0.14\columnwidth} l}
  \toprule
  ID & Design Guidelines & Engineering Experience & Literature Support \\
  \midrule
  DG1 &
  Provide stable and robust automation assistance that supports developers' workflows, including routine and repetitive tasks. & Sec. III.A.1.    Sec. III.A.3. 
    Sec. III.B.
   & \cite{wessel2018power, wessel2022bots,robe2022pair, liu2020understanding}
   \\ \addlinespace[0.3em]
    DG2 & Enable customization of the bot's behavior to align with project and ecosystem norms. & Sec. III.B. & \cite{wang2022specialized, liu2020understanding}
   \\ \addlinespace[0.3em]
    DG3 & Support timely updates of access permissions and dependencies to ensure security. & Sec. III.A.2. & \cite{wang2022specialized}
   \\ \addlinespace[0.3em]
    DG4 & Minimize identifiable information exposure in data pipelines to address privacy concerns.
   & Sec. III.B. & -
   \\ \addlinespace[0.3em]
   DG5 & Provide simple and controllable feedback during interactions with developers, including control over the type and frequency of interactions. & Sec. III.A.2. Sec. III.A.4. & \cite{wessel2022guidelines, lebeuf2018software}
   \\ \addlinespace[0.3em]
   DG6 & Promote the reuse of automation components and transparency in the bot's operation. & Sec. III.B. & \cite{erlenhov2020empirical, liu2020understanding, brown2019sorry}
   \\
  \bottomrule
\end{tabularx}
\end{table}

\subsection{Expectations for Future SE Bot Development}

Based on the experiences of elite developers with SE bots, they expressed their expectations for both functional and non-functional features in future bot development. Firstly, elite developers highlighted the need for future bots to provide technical values similar to those currently available. For instance, P2 suggested that the functionality of future bots should build on the existing automation features, ``\textit{improve the functionality you want to see, and sum up to the current bot}.'' Participants expressed a desire for additional technical functions, such as programming language support for code coverage (P3), Issue-Pullrequest linkage (P2), and integration with JIRA issue tracking (P4), among others.

Elite developers highlighted the significance of ease of use and adhering to the principles of open-source software development, alongside technical functionality, as crucial considerations for designing future SE bots. A participant highlighted that the availability of SE bots for free aligns with the core values of OSS development, ``...\textit{if you}[r project is] \textit{free and you don't have to pay. I think the payment model is, you have to pay if you want your repo to be private.} (P3)'' Similarly, another participant (P2) expressed a desire for open-source reproduced versions of SE bots to be available, as some large-scale organizations choose to withhold implementation details due to security and privacy concerns, 
Moreover, they also noted that these high-quality bots could serve as valuable resources for developers seeking to improve their workflows.

\section{Synthesis and Design Guidelines}
\subsection{Summarizing Interview Results}
 Through conducting interviews with elite developers, our research found that these individuals incorporate SE bots extensively into their daily development workflow. Specifically, SE bots are utilized by developers in both user- and system-based ways, providing automation for various aspects of their work. Elite developers have expressed that SE bots offer significant technical value in their engineering workflow, as well as support for their community, allowing them to focus on other tasks. However, interactivity and excessive notifications have emerged as two significant challenges that elite developers face while using bots. When considering their expectations for future SE bots, elite developers prioritize simplicity in design and use, the technical value of automating repetitive tasks, and following the OSS spirit.
 
\subsection{Synthesizing with Prior Studies}

To provide comprehensive design guidelines for such a list, this study employs three major sources to elicit the design guidelines~\cite[p.~2]{brown1999human}, evidence gathered through engineering experience, and evidence from previous empirical studies and predictions from existing literature of bots (see Tab. \ref{tab:dg}).

Table \ref{tab:dg} summarizes the evidence supporting the design guidelines (DGs) proposed in this study. The six DGs aim to facilitate developers' daily workflow, inform future bot development efforts, and advance software engineering practices. Following these guidelines can improve the awareness of elite developers' productivity, efficiency, and security, while reducing privacy risks.
By following these design guidelines, future SE bots may enhance elite developers' productivity, efficiency, and well-being while minimizing potential risks.

\section{Concluding Disscussion}
In conclusion, SE bots have become a crucial and widely adopted extension of daily engineering processes in many OSS organizations, serving as the primary interface for elite developers maintaining OSS repositories. They have had a significant impact on the way elite developers attend to community needs, making their daily engineering processes more efficient. However, despite their prevalence, challenges remain in integrating them into developers' workflows, such as disruptive notifications, limited interactivity with novices and the community, not aligning with the OSS spirit, and consequentially, limited customization options. 

This study proposes six design guidelines to provide direction for bot developers and enhance the support they offer to elite developers. Note that effective strategies for allocating development resources may vary depending on the types of bots being developed. For instance, bots integrated into PR workflows may benefit from transparency and security in their decision-making~\cite{wessel2022bots}. Additionally, bots designed to support communities may need to prioritize interactivity and leverage conversational AI to enhance their effectiveness~\cite{brown2020language}. By following these guidelines, future SE bots have the potential to improve elite developers' productivity and well-being.


Moving forward, further research is needed to explore the effectiveness of these guidelines in improving SE bots' performance and addressing the identified limitations and challenges. Additionally, there is a need for ongoing evaluation of SE bots to ensure they continue meeting elite developers' changing needs as their responsibilities evolve. SE bots have the potential to improve software engineering practices and help OSS projects stay sustainable and competitive in the rapidly evolving technological landscape. However, continued efforts are necessary to ensure that SE bots are developed and deployed responsibly and inclusively, considering their impact on elite developers and the OSS community as a whole.

\section*{Acknowledgements}
The authors would like to express their gratitude to the anonymous participants and reviewers. Z. Wang and D. Redmiles are supported in part by the UC Irvine, Donald Bren School of ICS. The interview study was classified as exempt by the UCI Institutional Review Board (IRB) office.

\bibliographystyle{IEEEtran}
\bibliography{IEEEabrv, citations}

\end{document}